\documentclass{amsart}
\title {Essential Inputs and Minimal Tree Automata}
\author{Ivo Damyanov and  Slavcho Shtrakov}
\date{}

\address{Dept. of Computer  Sciences, South-West University,
 Blagoevgrad}
\email{damianov@aix.swu.bg, shtrakov@aix.swu.bg}
\date{}
\usepackage{graphicx}
\usepackage{graphics}
\usepackage{latexsym}
\usepackage{latexsym}
\usepackage{latexsym}
\newtheorem{l00}{\bf Lemma}

\newtheorem{pr00}{\bf Proposition}
\newtheorem{t00}{\bf Theorem}
\newtheorem{e00}{\bf Example}
\newtheorem{d00}{\bf Definition}
\newtheorem{c00}{\bf Corollary}
\def\Pr{{\bf Proof.\ }}
\def\fbx{\hfill${}^{\rule{2mm}{2mm}}$}
\def\la{\leftarrow}
\def\ra{\rightarrow}

\def\F{{ F}}
\def\A{{ A}}

\relax
\begin{document}

\maketitle
\begin{abstract}
\noindent In the paper we continue studying essential inputs of
trees and automata initiated in \cite {[Dr2000]}. We distinguish
the behavior of the essential inputs of trees and essential
variables for discrete functions. Strongly essential inputs of
trees are introduced too. It is proved that if a tree and an
automaton have at least two essential inputs then they have at
least one strongly essential input. A minimization algorithm for
trees and automata is proposed. Various examples for application
in Computer Science are shown.

\end{abstract}

\maketitle
 \noindent {\it AMS, subject classification}: 03D05, 68Q70, 03D15, 06B25

\section{Introduction}
Tree automata are designed in context of circuit verification
and logic programming. In the 1970's some new results were obtained
concerning tree automata, as an important part of theoretical basis of
the computing
and programming. So, since the end of 1970's tree automata
have been used as powerful tools in program verification.
There are many results connecting properties of programs or
type systems or rewrite systems with automata.
\\
In the papers of S.Jablonsky \cite{Jab}, A.Salomaa \cite{Sa},
K.Chimev \cite{Ch} etc. the theory of essential variables for
discrete functions was developed. Some new interpretation for
essential, and strongly essential variables were introduced in \cite
{Da; 2001}.
\\
The concept of essential variables and separable sets of variables
has been introduced for terms in Universal algebra by K. Denecke
and Sl. Shtrakov \cite{[Sh-De 98]}. In \cite {[Dr2000]} the second
author of this paper initiate the investigation of the behavior or
essential input variables for tree automata and trees.

\section{Basic Definitions and Notations}
Let ${F}$ be any finite set, the elements of which are called $ operation\
symbols. $ Let $\tau:{{F}}\to N$ be a mapping into the non negative integers;
for $f\in{ F},$ the number $\tau(f)$ will denote the {\it arity } of the
operation symbol $f.$ The pair $(F,\tau)$ is called {\it type} or
{\it signature}.
 Often if it is obvious what the set ${F}$ is,
we will write "$ type\ \tau$". The set of symbols of arity
$p$ is denoted by ${ F}_p.$ Elements of arity $0,1,\ldots, p$
respectively are called {\it constants(nullary), unary,...,$p$-ary} symbols.
We assume that ${ F}_0\neq\emptyset.$
\begin{d00}\rm\label{d2}
Let $X_n=\{x_1,\ldots,x_n\}, n\geq 1,$ be a set of variables with
$X_n\cap{{F}}=\emptyset.$
The set $W_{\tau}(X_n)$ of {\it  $n-$ary terms (trees) of\ type} $\tau$
with\ variables\ from\ $ X_n $ is defined as the smallest
set for which:
\\
$(i)$\ $F_0 \cup X_n \subseteq W_{\tau}(X_n)$
\\
$(ii)$\ if $p\geq 1, f\in F_p $ and
$t_1,\ldots,t_p\in W_{\tau}(X_n)$ then $f(t_1,\ldots,t_p)\in W_{\tau}(X_n).$
\end{d00}
By $W_\tau(X)$ we denote the following set
$$W_\tau(X):=\cup_{n=1}^{\infty} W_\tau(X_n),$$
where $X=\{x_1,x_2,\ldots \}.$
If $X=\emptyset$ then $W_\tau(X)$ is also written $W_\tau.$ Terms in
$W_\tau$ are called {\it ground terms.}
\\
Let $t$ be a term. By $Var(t)$  the set of all variables from
$X$ which occur in $t$ is denoted. The elements of $Var(t)$ are  called
{\it input variables or inputs} for $t$.
 \\
Let $t$ be a term and suppose we are given a term $s_x$ for every $x\in X.$
The term denoted by $t(x\la s_x),$ is obtained by substituting in $t,$
simultaneously for every $x\in X,\quad s_x$ for each occurrence of $x.$
\\
If  $t,s_x\in W_{\tau}(X)$  then $t(x\la s_x)\in W_{\tau}(X).$
\\
Any subset $L$ of $W_{\tau}(X)$ is called {\it term-language}
or {\it tree-language.}
\\
Let $t$ be a term of type $\tau.$ We define the $depth$ of $t$
inductively as follows:\\
$(i)$ if $t\in X\cup F_0$ then $Depth(t)=0;$\\
$(ii)$ if $t=f(t_1,\ldots,t_{n})$ then
$Depth(t)=max \{Depth(t_1),\ldots,Depth(t_{n})\} +1.$
\\
Let $N$ be the set of natural numbers and $N^*$ be the set of
 finite strings over $N.$  The set $N^*$ is naturally ordered by
$\forall \overline n, \overline m \in N^* \quad \overline n\preceq
\overline m \iff \overline n$\ is a prefix of $\overline m.$
\\
A term $t\in W_{\tau}(X)$ may be viewed as a finite ordered tree,
the leaves of which are labelled with variables or constant
symbols and the internal nodes are labelled with operation symbols
of positive arity, with out-degree equal to the arity of the
label, i.e. a term $t\in W_{\tau}(X)$ can also be defined as a
partial function $t:N^*\to {{F}}\cup X$ with domain $Pos(t)$
satisfying the following properties:
\\
$(i)$\quad $Pos(t)$ is nonempty and prefix-closed;
\\
$(ii)$\quad For each $p\in Pos(t)$, if $t(p)\in { F}_n,$
$n\geq 1$
then $\{i|pi\in Pos(t)\}=
\{1,\ldots,n\};$
\\
$(iii)$\quad For each $p\in Pos(t)$, if $t(p)\in X\cup F_0$
then $\{i|pi\in Pos(t)\}=\emptyset.$
\\
\par
The elements of $Pos(t)$ are called  {\it positions.} A {\it frontier position} is a position
$p$ such that
$\forall \alpha\in N,\quad p\alpha\notin Pos(t).$
 Each position $p$ in $t$ with $t(p)\in X$ is called {\it variable position}
and if $t(p)\in F_0$ it is
called {\it constant position.}
\par
A {\it subterm} $t|_p$ of a term $t\in W_{\tau}(X)$ at position $p$ is
defined as follows:
\\
$(i)$\quad $Pos(t|_p)=\{i|pi\in Pos(t)\};$
\\
$(ii)$\quad $\forall j\in Pos(t|_p),\quad t|_p(j)=t(pj).$
\\
The subtrees at the frontier positions for $t$ are called {\it
inputs} of $t.$
\par
 By $t[u]_p$ we  denote the term obtained by replacing the subterm $t|_p$ in $t$
 by $u.$
\\
 We write $Head(t)=f$ if and only if $t(\varepsilon)=f$, where
$\varepsilon$ is the empty string in $N^*,$
i.e. $f$ is the {\it
root symbol} of $t.$
\\
Thus we define a partial order relation in the set of all terms $W_{\tau}(X).$
We denote by $\unlhd$ the subterm ordering, i.e. we write $t\unlhd t'$ if
there is a position $p$ for $t'$ such that $t=t'|_p$ and one says that
$t$ is a subterm of $t'.$ We write $t\lhd t'$ if $t\unlhd t'$ and
$t\neq t'.$
\\
A chain of subterms $Ch:=t_{p_1}\lhd t_{p_2}\lhd \ldots \lhd t_{p_k}$ is called {\it strong}
if for all $j\in \{1,\ldots,k-1\}$
there does not exist a term $s$ such that
$t_{p_j}\lhd s\lhd  t_{p_{j+1}}.$
\vspace{-.2cm}

\section{Finite Tree Automata and Essential Variables}
\begin{d00}\rm\label{d5}
A  {\it finite tree automaton}  is a tuple
$\A=\langle Q,\F, Q_f,\Delta\rangle$ where:
\\
- $Q$ is a finite set of states;
\\
- $Q_f\subseteq Q$ is a set of final states;
\\
- $\Delta$ is a set of transition rules i.e. if
$$\F= \F_0\cup\F_1\cup\ldots\cup\F_n\quad
{\mbox{\rm then}}\quad \Delta=\{\Delta_0,\Delta_1,\ldots,\Delta_n\},$$
where  $\Delta_i$
are  mappings
$\Delta_0:F_0\ra Q,$
and
$\Delta_i:\F_i\times Q^i\ra Q,$ for $ i=1,\ldots,n.$
\end{d00}
\noindent
We will suppose that $\A$ is complete i.e. the $\Delta$'s are total mappings on their
domains.
\\
Let $Y\subseteq X$ be a set of variables and
$\gamma:Y\ra \F_0$
be a function which assigns  nullary operation symbols (constants) to each input
variable from $Y.$
 The function $\gamma$ is called
{\it assignment on the set of inputs $Y$} and the set of such assignments
 will be denoted by
$Ass(Y,\F_0).$
\\
Let $t\in W_{\tau}(X),$ $\gamma\in Ass(Y,\F_0)$ and
$Y=\{x_1,\ldots,x_m\}.$ By $\gamma(t)$ the term
$\gamma(t)=t(x_1\la \gamma(x_1),\ldots,x_m\la \gamma(x_m))$
will be denoted.
 \\
So, each assignment $\gamma\in Ass(Y,\F_0)$ can be extended to
a mapping defined on the set
$W_{\tau}(X)$ of all terms.
\\
Let $t\in W_{\tau}(X),$ and $ \gamma\in Ass(X,\F_0).$
The automaton
$\A=\langle Q,\F, Q_f,\Delta\rangle$
 runs over $t$ and $\gamma.$ It starts at
leaves of $t$ and moves downwards, associating along a run a
resulting state with each subterm inductively:
\\
$(i)$\quad If $Depth(t)=0$ then the automaton $\A$ associates with $t$ the
state $q\in Q,$ where
$$q=\left\{\begin{array}{ll}
       \Delta_0(\gamma(x_i))  & \quad \mbox{\rm if}\quad t=x_i\in X;\\
       \Delta_0(f_0)  & \quad \mbox{\rm if}\quad t=f_0\in\F_0.
        \end{array}
        \right.$$
$(ii)$\quad Let $Depth(t)\geq 1.$
If $t=f(t_1,\ldots,t_n)$ and the states  $q_1,\ldots,q_n$ have been
associated with the subterms(subtrees) $t_1,\ldots,t_n$ then with $t$
the automaton $\A$ associates the state $q,$ according to
$q=\Delta_n(f,q_1,\ldots,q_n).$
\\
The automaton runs only over ground terms and each
assignment from $Ass(X,F_0)$ transforms any tree as a ground term.
\\
The initial states are the states associated with the leaves of the tree
as for terms with depth equals to 0 i.e. as in the case $(i).$
\\
A term $t\ ,t\in W_{\tau}(X)$ is accepted by a  tree automaton
$\A=\langle Q,\F, Q_f,\Delta\rangle$
if there exists an assignment $\gamma$ such that when running over $t$ and
$\gamma$ the automaton $\A$ associates with $t$ a final state  $q\in Q_f.$
\\
When $\A$ associates the state $q$ with a subterm $s,$ we will write
$ \A(\gamma,s)=q.$
\\
Let $t\in W_{\tau}(X)$ be a term and $\A$ be a  tree automaton which accepts
$t.$ In this case one says that $\A$ {\it recognizes} $t$ or $t$ is
{\it recognizable} by $\A.$ The set of all by $\A$
 recognizable terms is called {\it tree-language} recognized by $\A$ and will be denoted
by $L(\A).$
\begin{d00}\rm\label{d6}
Let $t\in W_{\tau}(X)$ and let $A$
be a tree automaton. An input variable $x_i\in Var(t)$ is called
{\it essential} for the pair $(t,{ A})$
if there exist  two
assignments $\gamma_1, \gamma_2\in Ass(X,{ F}_0)$ such that
$$\gamma_1(x_i)\neq \gamma_2(x_i),\quad
\forall x_j\in X,\ j\neq i \quad \gamma_1(x_j)=\gamma_2(x_j)$$
 with
${ A}(\gamma_1,t)\neq { A}(\gamma_2,t)$
i.e. ${ A}$ stops in
different states when running over $t$ with  $\gamma_1$ and  with
$\gamma_2.$
\end{d00}
The set of all essential inputs for $(t,{ A})$ is denoted by
$Ess(t,{ A}).$ The input variables from
$Var(t)\setminus Ess(t,{ A})$
are called {\it fictive } for  $(t,{ A}).$

\begin{e00}\label{e1}\rm
Let ${A}=\langle Q,{ F}, Q_f,\Delta\rangle$  with
\\
${ F}_0=\{0,1\}$, ${ F}_1=\{f_1\}$, ${ F}_2=\{g_1,g_2\},$
$Q=\{q_0,q_1\}$, $Q_f=\{q_1\}$,
\\
$\Delta_0(0)=q_0$, $\Delta_0(1)=q_1$,
$\Delta_1(f_1,q_0)=q_1$,
$\Delta_1(f_1,q_1)=q_0$,
\\
$\Delta_2(g_1,q_0,q_1)=
\Delta_2(g_1,q_1,q_0)=
\Delta_2(g_1,q_1,q_1)=q_1$,
$\Delta_2(g_1,q_0,q_0)=q_0$,
\\
$\Delta_2(g_2,q_0,q_0)=\Delta_2(g_2,q_0,q_1)=
\Delta_2(g_2,q_1,q_0)=q_0$,
$\Delta_2(g_2,q_1,q_1)=q_1.$
\\
Let us consider the term
$t=g_2(g_1(f_1(x_2),x_1),x_1).$
\\
The tree of the term $t$ is given on the Figure \ref{f1}:

\begin{figure}
  \includegraphics[width=10cm]{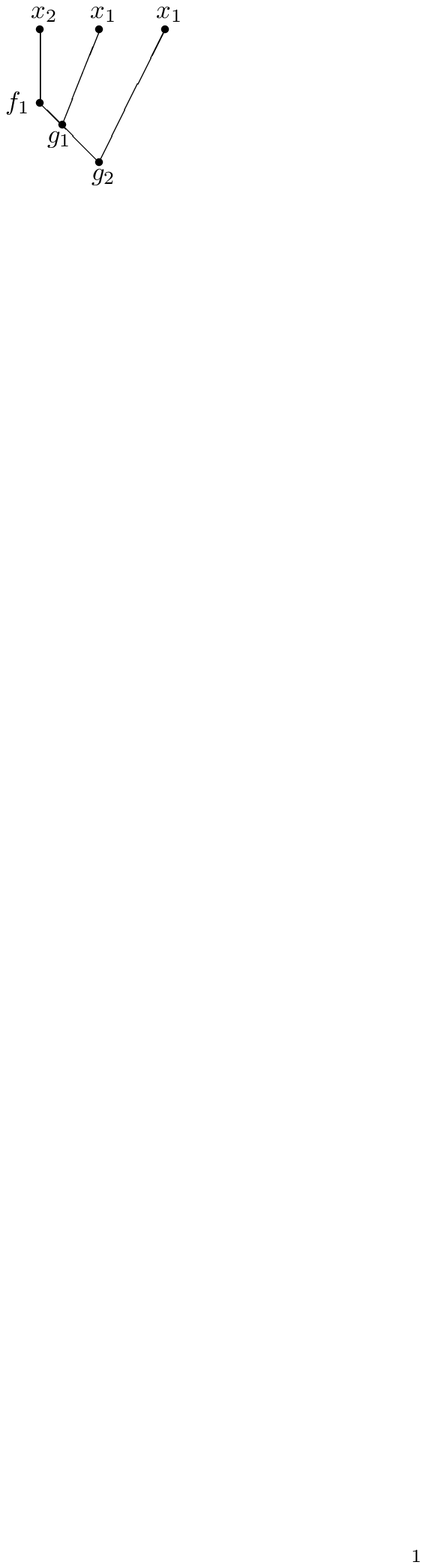}\\
  \caption{~~~}\label{f1}
\end{figure}

The set of positions for $t$ is:
\\
$Pos(t)=\{\varepsilon,1,11,111,12,2\}$
and the corresponding subtrees to these positions are:
$t|_{1}=g_1(f_1(x_2),x_1),$\
$t|_{11}=f_1(x_2),$\
$t|_{12}=x_1,$\
$t|_{111}=x_2,$\
$t|_{2}=x_1.$\
\\
There are four possible assignments and exactly three strong chains of subterms
which connect the leaves of $t$ and  the root of $t.$
\\
 It is easy to see that
$x_2 \in Ess(t|_{1},{ A}),$ and $x_2 \in Ess(t|_{11},{ A}),$ but
$ x_2\notin Ess(t,{ A}).$
\end{e00}

When investigating the finite valued functions with respect to their
essential variables and their subfunctions a remarkable result says
that \cite{Ch}: {\em if a variable $x_i$ is essential for a
subfunction $f_1$ of $f$ then there is a chain
$$f_1 \prec f_2\prec \dots \prec f_n=f,$$
such that $x_i$ is essential for $f_j,$ where $j=1, 2,\dots, n$ and $h\prec g$ means that $h$
is a subfunction of $g.$}
\\
This result for trees and automata is not held.
\\
Consider the subtree $t_1 = g_1 (f_1(x_2), x_1)$ of the tree $t$ given in the
Example \ref{e1}. It is
easy to see that $x_2 \in Ess (t_1, A)$  but $x_2 \not\in Ess (t,A).$ In \cite {[Dr2000]} the
following theorem is proved.

\begin{t00}\label{t1}\rm
If $x_i\in Ess(t,{ A})$ then there exists a strong
chain
$x_i=t_1\lhd t_2\lhd \ldots \lhd t_k\unlhd t$
such that $x_i\in Ess(t_j,{ A})$ for $j=1,\ldots,k.$ \fbx
\end{t00}

\begin {pr00}\label{prp1}
 $\forall \gamma\in Ass(X,{ F}_0)\quad
 { A}(\gamma,t')={ A}(\gamma,t) $
then
$Ess(t,{ A})=Ess(t',{ A}).$
\end {pr00}

Another important result for finite valued functions concerns
strongly essential variables which we will prove for trees and
 automata, which is the aim of the next section.

\section {Strongly Essential Inputs}

\begin{d00}\rm\label{dse}
Let $t\in W_{\tau}(X)$ and let $A$ be a tree automaton and
$M \subseteq Ess (t, A)\ (M \neq\emptyset).$
An input variable $x_i\in M$ is called
{\it strongly essential} for the pair $(t,{ A})$ with respect to set $M$
if there exist value $f_0$ for the input $x_i$ such that
$M\setminus \{x_i\} \subseteq Ess (t(x_i\la \F_0), A).$
\end{d00}

\begin{l00}\rm\label{lma}
Let $Ess (t,A) = Y_1 \cup Y_2,$ $Y_i \neq \emptyset,$ $Y_1\cap Y_2= \emptyset.$
If there is an assignment
$\gamma \in Ass (Y_2, \F_0)$ such that $Y_1 = Ess (\gamma(t), A)$
then there is an input $x_i\in Y_2$ which
is strongly essential for $t$ with respect to $A.$
\end {l00}

\Pr

At first let $Ess (t,A) = \{x_1, x_2\}.$
Clearly both $x_1$ and $x_2$ are strongly essential
with $Y_1 = \{x_1\},$ and $Y_2=\{x_2\}.$

Suppose that for each $s\in W_\tau (X)$ with
 $Ess (s,A) = Y_1 \cup Y_2,$ $Y_i \neq \emptyset,$
$Y_1 \cap Y_2 = \emptyset,$ $|Y_2| \leq l$ and there
 is an assignment $\gamma\in Ass (Y_2,\F_0)$
such that $Y_1 = Ess (\gamma(s), A)$ then there exists a
strongly essential input $x_i\in Y_2$
of $s$ with respect to $A.$

Let us consider a tree $t$ with
$Ess (t,A) = Y_1 \cup Y_2,$ $Y_1 \cap Y_2 = \emptyset,$ $|Y_2|=l+1$
and there is an assignment $\gamma\in Ass (Y_2,\F_0)$ such that
$Y_1 = Ess (\gamma(t), A).$

Suppose that $Y_2 = \{x_{m+1}, ..., x_{m+l+1}\}.$
Let $t_1 = t(x_{m+1} \la \gamma (x_{m+1})).$

If $Y_2\setminus \{x_{m+1}\} \subset Ess (t_1, A)$
then clearly $x_{m+1}$ is  strongly essential
input for $t$ with respect to $A.$

Consider the case $Y_2\setminus\{x_{m+1}\} \not\subset Ess (t_1,A)$ and let
$x_j\in \big( Y_2\setminus \{x_{m+1}\}\big ) \setminus \left ( Ess (t_1, A)\right).$

This means that for each $f_0 \in \F_0$ and for each
$\gamma_1 \in Ass (Y_2\setminus \{x_{m+1}, x_j\})$
$A(\gamma_1,t_1) = A(\gamma_1, t_2)$ where $t_2 = t_1 (x_j \la f_0).$
Let $f_0'$ be such that $x_{m+1} \in Ess (t_3, A)$ where $t_3 = t(x_j\la f_0').$

It is clear that
$$Y_1 \subset Ess (t_3, A).$$

Let us set $Y_3 = Y_2 \setminus Ess (t_3, A).$ Obviously $Y_3 \neq \emptyset$ (note that
$x_j \in Y_3$). On the other hand $x_{m+1} \not \in Y_3$ and $Y_3 \subset Y_2.$
Clearly $|Y_3| \leq l.$ Let us set
$Y'_1 = Y_1 \cup (Y_2\setminus Y_3)$ and $Y'_2 = Y_3.$
By $Y_3 \cap Ess (t_3, A) = \emptyset$ and $Ess (t_3, A) = Y'_1$ it follows that
there is at least one assignment $\gamma' \in Ass (Y'_2, \F_0)$ such that
$$Y'_1 = Ess (\gamma', t).$$
By the inductive assumption it follows that
 there is an input $x_r\in Y_3$ which is strongly
  essential input for $t$ with respect to $A.$
\fbx
\begin{t00}\label{tse}\rm
Let $t\in W_{\tau}(X)$ and let $A$ be a tree automaton. If $|Ess (t,A)| \geq 2$ then
there is at least one strongly essential input of $t$ with respect to $A.$
\end{t00}
\Pr

Let$Ess (t,A) = \{x_1, \ldots, x_n\}.$ By $x_1 \in Ess (t, A)$ it follows that there is an
assignment $\gamma \in (Y_2, \F_0)$ with $Y_1 = Ess (\gamma (t), A)$ where $Y_1 = \{x_1\} $
and $Y_2 = \{x_2, \ldots, x_n\}.$

>From this the lemma \ref {lma} implies the proof of the theorem.
\fbx

\begin {c00}
Let $t\in W_{\tau}(X)$ and let $A$ be a tree automaton. If $|Ess (t,A)| \geq 2$ then
there is at least two strongly essential input of $t$ with respect to $A.$
\end {c00}

\section{Minimal Tree Automata}
In this section we consider minimization algorithms for trees and automata.

Proposition \ref {prp1} shows that if $t_1 \lhd t_2 \lhd t$ and
$\forall \gamma \in Ass (X, F_0)$ $ A(\gamma, t_1) = A(\gamma, t_2)$ then
$A(\gamma, t) = A(\gamma', t')$ for all $\gamma'\in Ass(X, F_0)$ and $$t' = t(t_2\la t_1).$$
Clearly if $t_1$ is a proper subtree of $t_2$ then $t'$ is a tree obtained from $t$ with
a reduction of the nodes i.e. $t'$ is more simple than $t.$

Another reduction can be obtained by removing of all non essential inputs of $t.$

These two operations (replacing $t_1$ and $t_2$ and removing the fictive inputs)
are used to reach minimal trees w.r.t. an automaton $A.$

\begin{d00}
A tree $t$ and an automaton $A$ are minimal if there
 are not any operations for reduction of $t.$

Clearly, the algorithm to find out minimal tree,
 automaton consist of applying all possible
reductions on the tree w.r.t. the automaton.
\end {d00}

 \section{Applications}

 Tree automata were designed in the context of circuit verification
 and logic programming. Becoming an important part of theoretical basis of
 the computing and programming, tree automata have been used as powerful tools
 in program verification.  In present computer technologies there are many
 examples where we can find the underlying tree automata.
 \medskip \\
 {\bf GUI}\medskip\\
 Powerful and intelligent Graphical
 User Interface (GUI) interacting with menus, dialogs, icons, etc. have
 hierarchical structure. The interactions with an element of the GUI
 reflect on the whole GUI. Each object send messages to the parent
 object on any action. The process for message passing between GUI
 objects is organized as automaton working over tree.
 \medskip \\
 {\bf XML}
 \medskip \\
 Databases as a concept for storing information
 is one of the major parts of the computer technology. Several main
 types of databases were affirmed. Now the dominating relational
 databases are going to be replaced by the well known hierarchical
 databases, using the XML technology. In the XML documents the nodes
 are divided in two types - nodes and attributes. The attributes are
 leaves and the nodes are the inner nodes of the tree. One to one
 mapping between XML document and tree exist. There are several
 manipulations with XML documents that the XML parser process as a
 Tree Automata, i.e. XSL translation, work with the DOM, validation
 with DTD.
\medskip \\
{\bf OOP}
\medskip \\
 In Object Oriented Languages such as C++ and Java, a user defined data
 type, a 'class', is introduced. Classes of objects can be put into
 hierarchy. Each class may contain fields that are variables or
 methods. Class fields may have different visibility. Again there is
 one to one mapping between class hierarchy and trees.
 One class can derive from another in different ways (using visibility
 modificators) which reflect on the visibility of the inherited fields.
 During the syntax checking of the program the translator works as a tree
 automata calculating the visibility of the class fields.

\label{endpag}

\begin{thebibliography}{10}

\bibitem{Com} H. Comon, M. Dauchet,
R. Gilleron, F. Jacquemard, D. Lugiez, S. Tison, M. Tommasi,{\it Tree Automata, Techniques
and Applications,} 1999,
http://www.grappa.univ-lille3.fr/tata/
\bibitem{Ch} K. Chimev, {\it Separable Sets of Arguments of
Functions}, MTA SzTAKI Tanulmanyok, 180/1986, 173 pp.
\bibitem{Den3}  K.Denecke, D.Lau, R.P\"oschel, D.Schweigert {\it
Hyperidentities, Hyperequational Classes and Clone Congruences,}
General Algebra 7, Verlag H\"older-Pichler-Tempsky, Wien 1991,
Verlag B.G. Teubner Stuttgart, pp.97-118
\bibitem{Da; 2001}  I. Damyanov {\it
On Some Properties of Variables in Reed-Muller Decompositions}
Mathematics and Education in Mathematics, Proceedings of $30^{th}$
Spring Conference of the Union of Bulgarian Mathematicians, Borovets, 2001, pp.258-262
\bibitem{Ges}  F. G\'ecseg, M. Steinby, {\it Tree Automata},
Akad\'emiai Kiad\'o, Budapest 1984
\bibitem{Gr} G. Gratzer,{\it General Lattice Theory}, Akad.-Verlag,
Berlin,1978
\bibitem{Jab} S. Jablonsky, {\it Functional Constructions in $k-$Valued Logic} (in Russian), Math. Institute V. Steklov, v.51, 1958, 5-142.
\bibitem{Mal} A. Mal'cev,{\it Algebraic Systems},(in Russian),
Nauka, Moscow,1970
\bibitem{Sa} A. Salomaa,{\it On Essential Variables of Functions,
Especially in the Algebra of Logic}, Ann.Acad.Sci.Finn., ser.A,333(1963),
1-11
\bibitem{[Dr2000]}Sl. Shtrakov, {\it Tree Automata and Essential Input Variables},
Contributions to General Algebra 13, Verlag Johannes Heyn,
Klagenfurt, 2000
\bibitem{Ros} I.G.Rosenberg,{\it \"Uber die funktionale Vollst\"andigkeit in den
mehrwertigen Logiken.} Roz. Ces.Akad.
ved, 80(1970), 3-93
\bibitem{[Sh-De 98]} Sl.Shtrakov,\ K.Denecke, {\it Essential Variables
and Separable Sets in Universal Algebra}, 1998, Multiple-Valued Logic Journal
\end{thebibliography}
\end{document}